\RequirePackage{fix-cm}
\documentclass[smallextended]{svjour3}       
\smartqed  
\usepackage{graphicx}
 \newcommand{\citep}{\cite}
\begin{document}

\title{Probabilities for Solar Siblings}

\author{Valtonen, Mauri \and
       Bajkova, A. T.  \and
       Bobylev, V. V. \and
       Myll\"ari, A.}


\institute{M. Valtonen \at
          FINCA, Univ. of Turku, 21500 Piikki\"o, Finland \\
          Tel.: +358-2-3338215\\
          Fax: +358-2-3335070\\
          \email{mvaltonen2001@yahoo.com}       
          \and 
A. T. Bajkova \at Central (Pulkovo) Astronomical Observatory of RAS,
 65/1 Pulkovskoye Chaussee, St-Petersburg, 196140, Russia
 \and  
V. V. Bobylev \at Pulkovo Observatory and Sobolev Astronomical Institute, St-Petersburg State University,
 Bibliotechnaya pl. 2, St-Petersburg, 198504, Russia
 \and
 A. Myll\"ari
 \at St.George's Univ., Grenada}
\date{Received: date / Accepted: date}

\maketitle

\begin{abstract}
We have shown previously (Bobylev et al 2011) 
that some of the stars in the Solar neighborhood today may have originated in the same star cluster as the Sun, and could thus be called Solar Siblings. In this work we investigate the sensitivity of this result to Galactic models and to parameters of these models, and also extend the sample of orbits. There are a number of good candidates for the Sibling category, but due to the long period of orbit evolution since the break-up of the birth cluster of the Sun, one can only attach probabilities of membership. We find that up to $10\%$  (but more likely around $1\%$) of the members of the Sun's birth cluster could be still found within 100 pc from the Sun today. 
\end{abstract}

\section{Introduction}
\label{intro}
Stars are generally born in clusters. Therefore it is likely that our Sun was also a member of a star cluster some 4 billion years ago after it had condensed from an interstellar gas cloud (Adams 2010). By now the cluster has dispersed and its stars have spread out over a considerable region of the Galaxy (Portegies and Zwart 2009).

 There are several reasons to try to trace the other members of this cluster which we call Solar Siblings.  Just to mention one, the other members of the cluster must have had planetary systems of their own. Rocks flung out from one planetary system would sometimes have landed on a planet of another system. This could have lead to exchange of material between the systems, and even to exchange of life between the planets of the different systems  (Valtonen et al 2009). This may have important implications for the search of life in the Galaxy. If the origin of life is a very rare process, it is possible that life in our Galaxy is confined only to the planets of the Solar Siblings, outside our Solar System. When evidence of life in exoplanetary systems will be perhaps uncovered one day, it is important to determine if such systems belong to Solar Siblings, or if they are members of the general star field in our neighborhood. Life among Solar Siblings alone would speak for the rarity of the origin of life. In this paper we ask whether it is still possible to recognize the Solar Siblings after some twenty Galactic revolutions and plenty of unpredictable orbit evolution in the past 4 billion years.

One way to search for Solar Siblings is to find stars of the same age and metallicity as the Sun. The Sun has an exceptional metallicity for our neighborhood  (Wielen et al 1996), with only $25\%$ of otherwise similar stars having metallicity as high or higher than the solar value. Thus metallicity studies of the nearby stars may provide a good set of Solar Sibling candidates.  Batista and Fernandes (2012)
 emphasize the chemical similarities and age in the search of the solar neighborhood, and pick out three candidate stars.  Bobylev et al (2011)
 look at the orbital convergence of stars in the past in a model Galaxy with spiral arms and identify two Solar Sibling candidates. The star HD83432 is common to both lists.  Brown et al (2010)
 find 6 candidates using a smooth Galaxy model without spiral arms. One of the stars in their list is common with  Batista and Fernandes (2012). Ramirez et al. (2014) suggest another star, HD162826, to be a Solar Sibling from a study of 30 candidates. In this work we vary the spiral pattern of the Galaxy in order to see the sensitivity of the result to the parameters of the pattern. Rather than looking for particular Solar Sibling candidates, we ask how likely it is that there are still today true Siblings near to us.

Previous studies suggest that we may find as many as $10\%$ of the original Sun's birth cluster members within 100 pc of us (Portegies and Zwart 2009), while others (Mishurov and Acharova 2011) put the number as low as below $0.1\%$. It is an interesting question which number is closer to truth. There are still large uncertainties as to the exact model of our Galaxy (Antoja et al. 2009, Gerhard 2011) as well as to the secular evolution of our Galaxy (Sellwood 2013). Therefore we try to cover a representative range of Galactic models.

\section{Construction of orbits}
\label{sec:2}
We calculate the stellar orbits by solving the
following system of equations of motion based on a realistic model
of the Galactic gravitational potential (Fernandez et al 2008):
\begin{equation}
\ddot{\xi}=-\frac{\partial\Phi}{\partial\xi}-\Omega^2_0(R_0-\xi)-2\Omega_0\dot{\eta},
\end{equation}
 $$
\ddot{\eta}=-\frac{\partial\Phi}{\partial\eta}+\Omega^2_0\eta+2\Omega_0\dot{\xi},
 $$$$
\ddot{\zeta}=-\frac{\partial\Phi}{\partial\zeta},
 $$
where $\Phi$ is the Galactic gravitational potential; the
 $(\xi,\eta,\zeta)$ coordinate system with the center at the Sun rotates
around the Galactic center with a constant angular velocity
$\Omega_0,$ with the  $\xi,\eta,$ and $\zeta$ axes being directed
toward the Galactic center, in the direction of Galactic rotation,
and toward the Galactic North Pole, respectively; $R_0$ is the
Galactocentric distance of the Sun.

We use the Fellhauer et al (2006) Galactic potential model. The
system of equations~(1) is solved numerically by the fourth-order
Runge--Kutta method, with the time step of one million years.

In the Fellhauer et al (2006) model, the Galactocentric distance
of the Sun is taken to be $R_0 = 8.0$~kpc and the circular
velocity of the Sun around the Galactic center is $V_0 =
|\Omega_0|R_0 = 220$~km s$^{-1}$. The axisymmetric Galactic
potential is represented as the sum of three components---the
central bulge, the disk, and the halo:
\begin{equation}
\Phi=\Phi_{halo}+\Phi_{disk}+\Phi_{bulge}.
\end{equation}

In this model,

-- the halo is represented by a potential dependent on the
cylindrical Galactic coordinates $R$ and $Z$ as
$\Phi_{halo}(R,Z)=\nu_0^2\ln(1+R^2/d^2+Z^2/d^2)$, where
$\nu_0=134$ km s$^{-1}$ and $d=12$ kpc;

-- the disk is represented by the potential from Miyamoto and
Nagai (1975) as a function of the same coordinates:
$\Phi_{disk}(R,Z)=-GM_d(R^2+(b+(Z^2+c)^{1/2})^2)^{-1/2}$, where
the disk mass $M_{d}=9.3\cdot10^{10}$ $M_{\odot}$, $b=6.5$ kpc,
and $c=0.26$ kpc;

-- the bulge is represented by the potential from Hernquist (1990): $\Phi_{bulge}(R)=-GM_{b}/(R+a)$, where the bulge
$M_{b}=3.4\cdot10^{10}$ $M_{\odot}$ and $a=0.7$ kpc.

If the spiral density wave is taken into account (Lin and Shu 1964, Lin et al 1969), then the following term is added to the
right-hand side of Eq.~(2) (Fernandez et al 2008):
\begin{equation}
 \Phi_{sp} (R,\theta,t)= A\cos[m(\Omega_p t-\theta)+\chi(R)],
\end{equation}
where
 $$
 A= \frac{(R_0\Omega_0)^2 f_{r0} \tan i}{m},
 $$$$
 \chi(R)=- \frac{m}{\tan i} \ln\biggl(\frac{R}{R_0}\biggr)+\chi_\odot.
 $$
Here, $A$ is the amplitude of the spiral wave potential; $f_{r0}$
is the ratio of the radial component of the perturbation from the
spiral arms to the Galaxy's total attraction; $\Omega_p$ is the
pattern speed of the wave; $m$ is the number of spiral arms; $i$
is the arm pitch angle, $i<0$ for a winding pattern; $\chi$ is the
phase of the radial wave (the arm center then corresponds to
$\chi=0^\circ$); and $\chi_\odot$ is the Sun's phase in the spiral
wave.

The spiral wave parameters are not very well determined (a review of the
problem can be found in Fernandez et al (2001) and Gerhard (2011)). The simplest model of a two-armed spiral pattern is
commonly used, although, as analysis of the spatial distribution
of young Galactic objects (young stars, star-forming regions, or
hydrogen clouds) shows, both three- and four-armed patterns are
possible (Russeil 2003, Englmaier et al 2008, Hou et al 2009, Bobylev and Bajkova 2014).
More complex models are also known, for example, the kinematic
model by Lepine et al (2001) that combines two- and four-armed
spiral patterns rigidly rotating with an angular velocity close to 
$\Omega_0$. Note also the spiral ring Galactic model of Melnik and Rautiainen (2009) that includes two outer rings elongated
perpendicular and parallel to the bar, an inner ring elongated
parallel to the bar, and two small fragments of spiral arms. As
applied to the Galaxy, the theories of nonstationary spiral waves
with a fairly short stationarity time (several 100 Myr), a
variable rotation rate, and a variable number of arms are also
considered (Sellwood and Binney 2002, Baba et al 2009). The
currently available data do not yet allow any one of the listed models
to be unequivocally chosen. Therefore, here we apply the models of
both stationary and non-stationary spiral pattern with different numbers of spiral arms, but disregard the influence of the bar.

We choose
$\Omega_p$ from the range $8-35$ km s$^{-1}$ kpc$^{-1}$ (Popova and Loktin 2005, Naoz and Shaviv 2007, Gerhard 2011). The pitch angle $i$ is known relatively well and is between
$-5^\circ$ and $-7^\circ$ for the two-armed spiral pattern and between $-10^\circ$ and $-14^\circ$ for the four-armed spiral pattern.
The amplitudes of the velocities of the perturbation from the
spiral density wave are $5-10$ km s$^{-1}$ (Mishurov and Zenina 1999, Fernandez et al 2001, Bobylev and Bajkova 2010) in both
tangential and radial directions.

The Sun's phase in the wave $\chi_\odot$ has a rather
large uncertainty. For example, according to Fernandez et al (2001), this angle lies within the range $284^\circ-380^\circ$.
Having analyzed the kinematics of open star clusters,  Bobylev et al (2008) and
Bobylev and Bajkova (2010) find $\chi_\odot$ close to
$-120^\circ$, while the data on masers yield an estimate of
$\chi_\odot = -130^\circ\pm10^\circ$ (Bobylev and Bajkova 2010); here,
the minus sign implies that the phase angle is measured from the
center of the Carina--Sagittarius spiral arm.

According to the classical approach in the linear density-wave
theory (Yuan 1969), the ratio $f_{r0}$ lies within the range
0.04--0.07 and the most probable value is $f_{r0} = 0.05$. The
upper limit $f_{r0} = 0.07$ is determined by the velocity
dispersion of young objects observed in the Galaxy (at $f_{r0} =
0.07,$ the dispersion must reach 25~km s$^{-1}$, which exceeds a
typical observed value of 10--15~km s$^{-1}$). We also study $f_{r0}=0.1$, see Antoja et al. (2009) and references therein.

We adopted the following optimum parameter values: for the two-armed spiral pattern
$(m=2),$ the basic pattern speed of the spiral wave $\Omega_p=20$
km s$^{-1}$ kpc$^{-1}$, the pitch angle $i=-5^\circ,$ the maximum
ratio $f_{r0}=0.05,$ and the Sun's phase in the wave $\chi_\odot =
-120^\circ$. We also took into account the Sun's displacement from
the Galactic plane $Z_\odot=17$~pc (Joshi 2007) and used the
present-day peculiar velocity of the Sun relative to the local
standard of rest $(U_\odot,V_\odot,W_\odot)_{LSR}=(10,11,7)$~km
s$^{-1}$ (Sch\"onrich et al 2010, Bobylev and Bajkova 2010). 

In addition, we studied the models of Sellwood (2010) and Gerhard (2011) with the following parameters: 

1 - model of Sellwood ($m=4$, $\Omega_p=18$ km s$^{-1}$ kpc$^{-1}$), 2 - model of Gerhard ($m=4$, 
$\Omega_p=25$ km s$^{-1}$ kpc$^{-1}$), 3 - model of Sellwood ($m=2$, $\Omega_p=8.1$ km s$^{-1}$ kpc$^{-1}$).

In the case of four-armed model ($m=4$) the pitch angle is taken as $i=-13^\circ$. Motivations for this value of pitch angle can be found 
in Bobylev and  Bajkova (2014).
From this and a kinematic analysis made earlier we find that the most 
probable number of spiral arms is $m=4$. In the case of two-armed Sellwood's model we took $i=-6^\circ$.

Our initial condition is such that one thousand stars are placed uniformly inside a disk of radius 10 pc with a velocity dispersion of 1 km/s in the two coordinates of the Galactic plane. They represent the birth cluster of the Sun in this two-dimensional simulation. Then their orbits are calculated forward for 4 billion years in a given Galactic potential, with the intermediate result noted at 2 billion years of integration. The pattern speed of the wave is varied from one calculation to another: its values are 8.1, 10, 15, 18, 20 or 25 km/s/kpc. Also we study the case of no spiral pattern, as well as the birth cluster being in exact corotation with the spiral pattern ($\Omega_{p}=200/8$ km/s/kpc), and 5 km/s/kpc off the corotation. The relative force of the spiral component is 5 or 10$\%$, the pitch angle is $-5^\circ$ or $-6^\circ$ for two-armed cases and $-13^\circ$ for 4-armed cases. The phase of the Sun in the spiral wave is $-120^\circ$. The cases include the models of Sellwood (2010) (see also Hahn et al (2011)) with $m=2$ and $m=4$ spiral arms, as well as Gerhard (2011) favoured model with $m=4$. We take the spiral arm force component to be either constant or variable as shown in Figure 1.

\begin{figure}
  \includegraphics[width=0.75\textwidth]{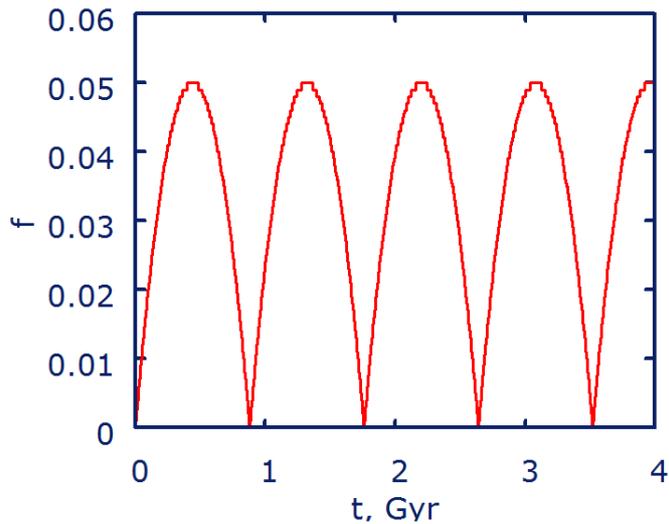}

\caption{The relative strength of the spiral pattern as a function of time in the model of variable spiral arms.}
\label{fig:1}       
\end{figure}

In the real Galaxy the density waves probably do not last more than about six revolutions, as they form and decay and reform over the twenty revolution periods  (Sellwood and Binney 2002, Sellwood 2012, Sellwood 2013). For this reason we construct models where the spiral arm strength varies in this manner. In the real Galaxy the evolution of the Galactic potential is likely to be more complex (Sellwood 2013); we use these models to illustrate the sensitivity of the nearby Solar Sibling numbers with respect to these parameters.
\section{Results}
\label{sec:3}
After the integration we look at the distribution of the Solar Sibling stars in the Galaxy (Figures 2(a) - 2(d)). Figures 2(a) and 2(b) compare the streams of stars after two different time spans, 2 Gyr and 4 Gyr, respectively. Results are shown both for the case of no spiral arms, and for constant spiral arms of two different pattern speeds. In Figure 2(c) we compare the effects of spiral arms which are in corotation with the initial cluster, or are off from the corotation by 5 km/s/kpc in either direction. In Figure 2(d) the spiral arm strength varies as illustrated in Figure 1.
\begin{figure}[H]
\begin{minipage}[h]{0.47\linewidth}
\center { \includegraphics[width=1\textwidth]{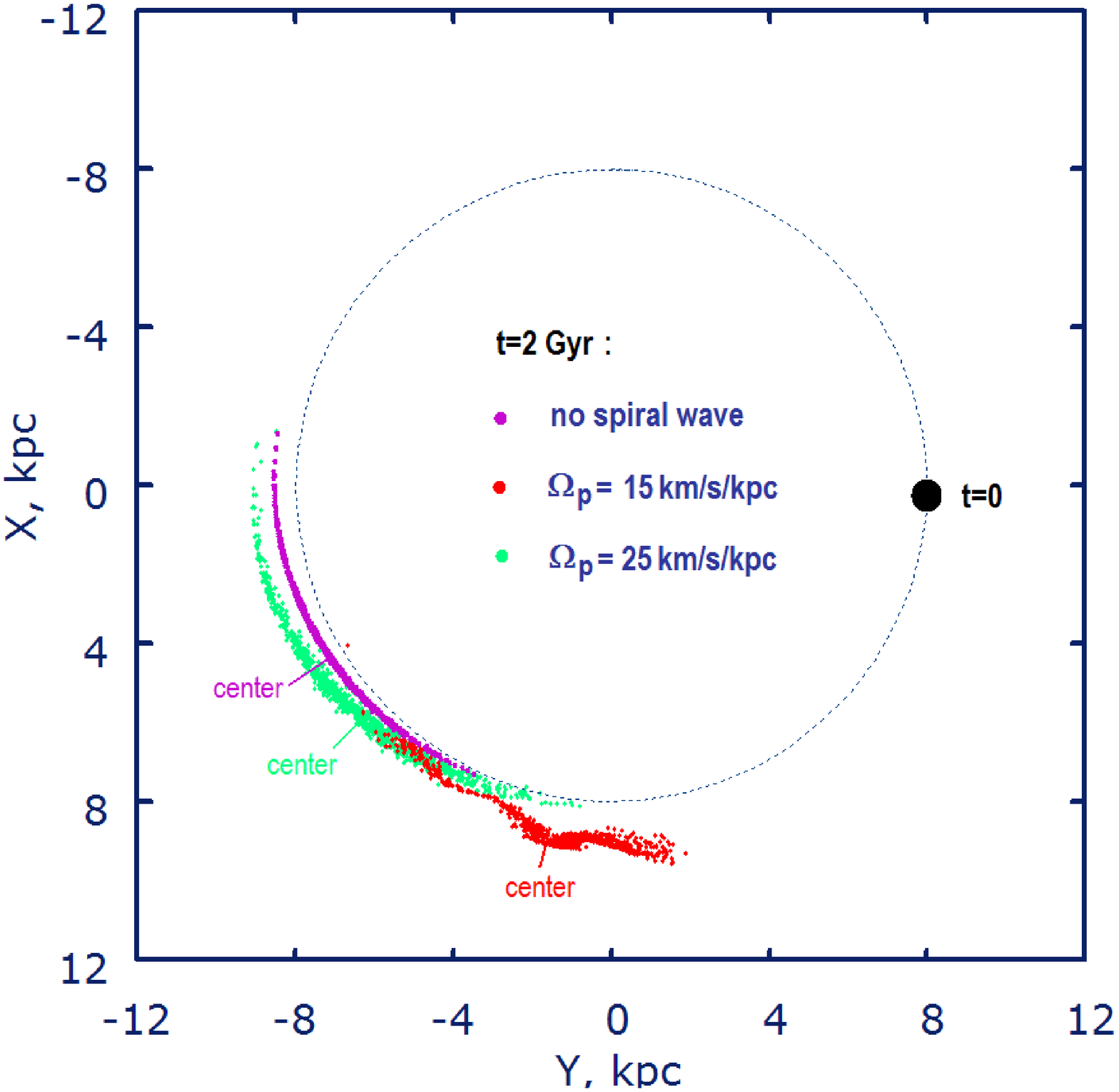} (a)}\\
\end{minipage}
\hfill
\begin{minipage}[h]{0.47\linewidth}
\center{ \includegraphics[width=1\textwidth]{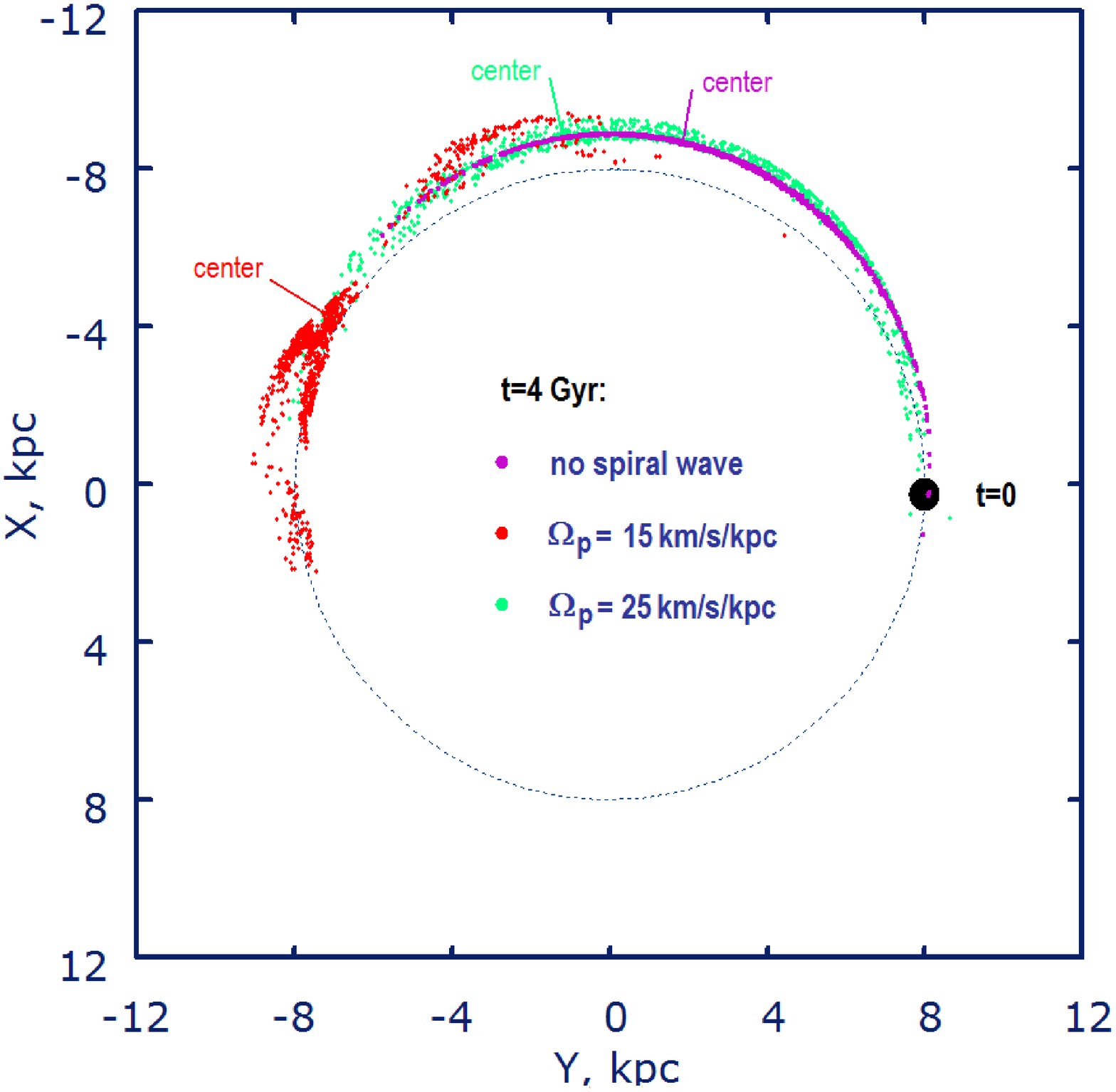} (b)}\\
\end{minipage}
\vfill
\begin{minipage}[h]{0.47\linewidth}
 \center{ \includegraphics[width=1\textwidth]{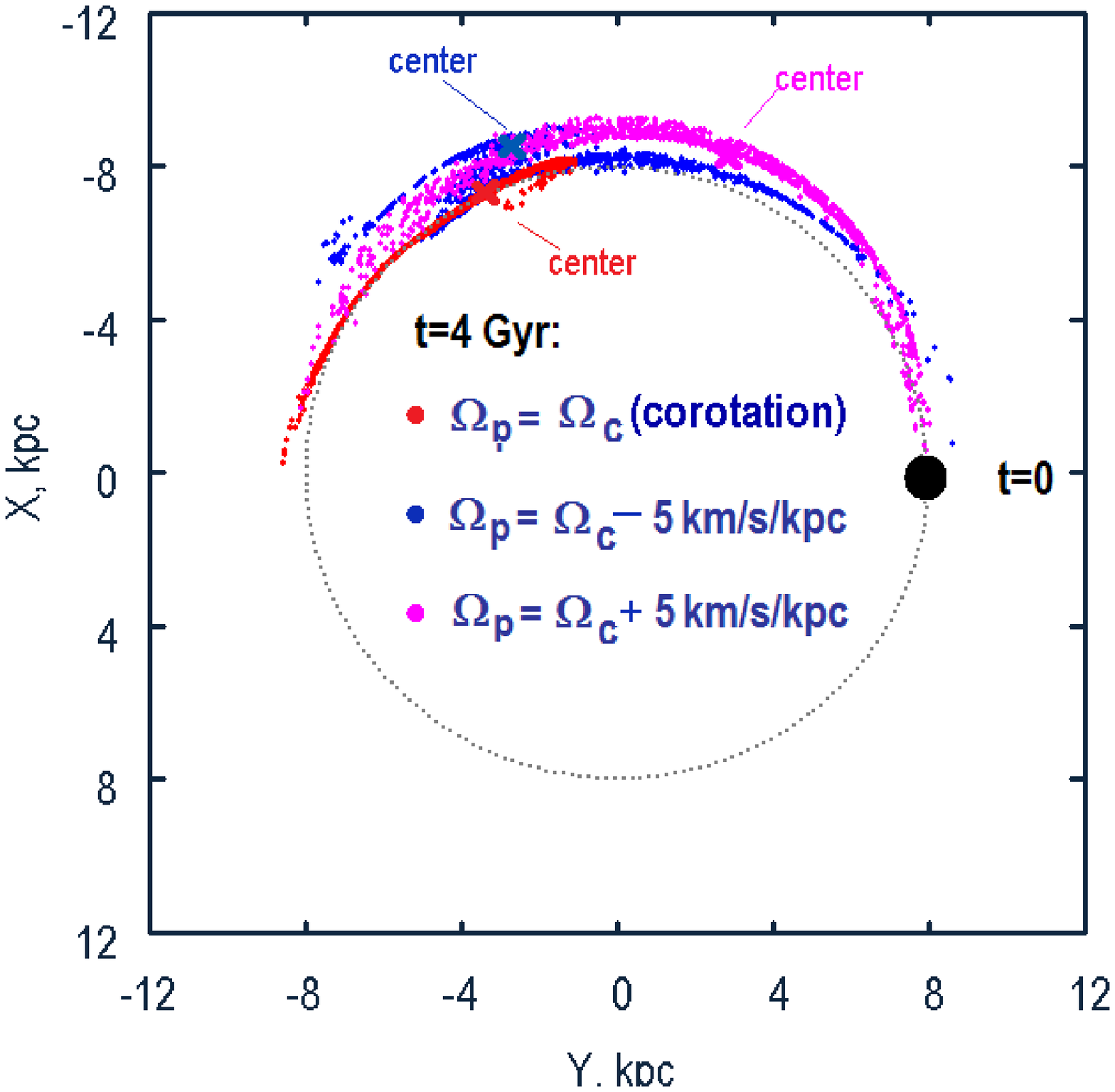} (c)} \\
\end{minipage}
\hfill
\begin{minipage}[h]{0.47\linewidth}
 \center{ \includegraphics[width=1\textwidth]{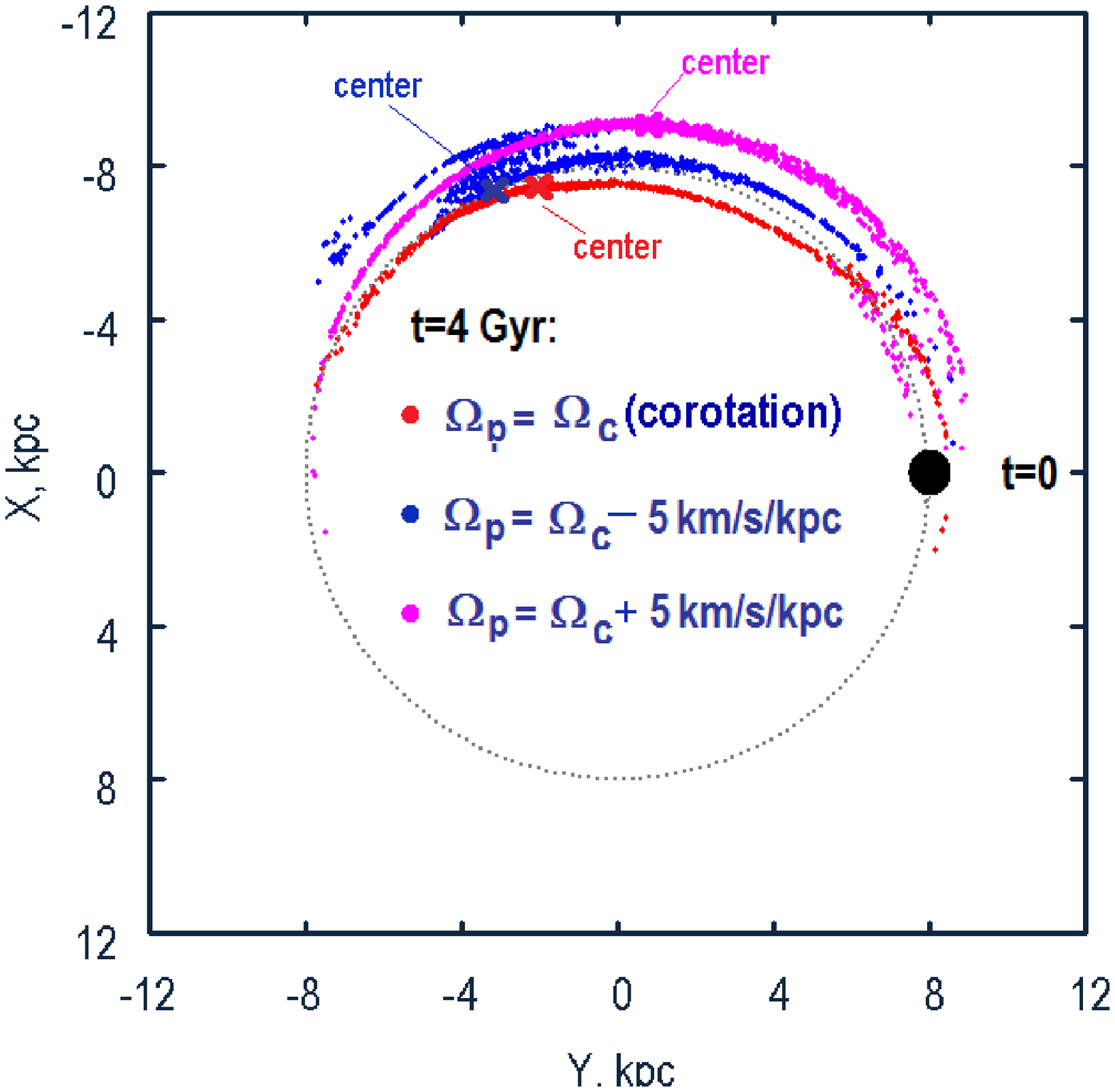} (d)}\\
\end{minipage}
\hfill
\caption{The 1000 stars originate from a cluster marked by a large dot at time $t=0$. After 2 Gyr panel (a) and after 4 Gyr (panels (b) - (d)) the stars have spread out in streams shown by small dots. Several cases of spiral pattern speed are shown as indicated in the figures. The center of each stream is also indicated. The circular orbit of the cluster center is depicted by a dotted line. In panel 2(d) the the spiral force component varies as shown in Figure 1, in other panels it is constant.}
\label{fig:2(a)}       
\end{figure}
%
%

%
To take a closer look at the differences between the different cases, we identify the center of each stream and study the $R=100$ pc, $R=50$ pc and $R=25$ pc surroundings around this center. This is not necessarily where the Sun would be in our model. However, it has been argued that the Sun was initially near the center of its birth cluster, and that it has not suffered close encounters with other stars (Adams 2010); thus it may still be not too far from the center of the stream. With this assumption, the number of stars within the specified circles represent the expected number of Solar Siblings today.

In Tables 1-4 we give the mean value of the number of the Solar Siblings for different models. The central values for $i$, $\Omega_p$ and $f_{r0}$ are listed in Column (1) of Tables 1-4 while for the central value for $\chi_\odot=-120^{\circ}$, if not otherwise stated. The parameters of the spiral wave $i$, $\Omega_p$, $f_{r0}$, $\chi_\odot$  were either fixed (Table 3) or varied over a wide range in accordance with a Gaussian distribution, using the mean values equal to the  nominal ones and the standard deviation $\sigma$ = nominal value / 2. The values beyond $3 \sigma$ were excluded. In Table 4 only the initial value of $f_{r0}$ was varied, around the central value of $f_{r0}=0.1$. In addition to a constant  $f_{r0}$, the time-varying  $f_{r0}$ was studied in all cases, using the stated maximum value.  We have carried out the Monte-Carlo statistical simulations using 100 random realizations for three models (Sellwood-1, Gerhard and Sellwood-2).

    The Solar Sibling mean values at R=100pc and after 4 billion years are typically $16\pm6$. We conclude that even though the spiral wave parameters may be uncertain to some degree, we expect Solar Siblings in all three models.

The maximum number of Solar Siblings (100 out of 1000 stars in the cluster) is found in the case of $\Omega_{p}=10$ km/s/kpc when $R=100$ pc at the time of 4 Gyr. This is rather a special case since the Galactic potential has refocussed orbits at 4 Gyr which were already more spread out at 2 Gyr. Even though it is an unlikely situation to be realized in the Galaxy, this possibility is also good to keep in mind. More typically, there are about ten Solar Siblings within the $R=100$ pc radius, and possibly none within the $R=25$ pc radius from the stream center.

Figures 3(a) and 3(b) plot the distance of a representative star from the cluster center in the constant spiral arm model and in the variable arm model, respectively, in the basic $f_{r0}=0.05$ model. The star makes repeated close approaches to the center. We may assume that the Sun is also in the central region of the cluster so that this represents also an approximation to the distance of the star from the Sun. We see that stars which make close approaches to the Sun do exist, but there is a certain amount of sensitivity with regard to the exact model. This is seen from the two lines which represent our standard variable spiral arm model, and another model where the spiral arm strength has been increased by $2\% $. Figure 3(c) shows a similar difference where the same initial orbit in the models of Sellwood (2010) with $m=4$ and $m=2$ is compared with Gerhard (2011) model of $m=4$, using the variable arm strength in each case. The diagram 3(d) for the constant arm strength shows qualitatively similar differences between the three models.

\section{Conclusions}
\label{sec:4}
We have found that the Portegies and Zwart (2009)
 estimate of 10 to 80 Solar Siblings within the 100 pc radius, for the same number of initial stars as in our calculation, is surprisingly good, even though this study neglected the effect of spiral arms. On the other hand, on the basis of Mishurov and Acharova (2011) we would expect mostly zeros in our tables, which is clearly not the case. Part of the explanation of the difference is likely to be in the amplitude of the spiral wave which is $10\%$ in the Mishurov and Acharova (2011) model while we use both  $5\%$ and $10\%$ in this work. However, even our $10\%$ models do give Solar Siblings, typically of the order of 10 (Table 2). Mishurov and Acharova (2011) also have randomly picked the $\Omega_{p}$ values in the range of 14.2 and 51.8 km/s/kpc which is another difference in the Galaxy model as compared with ours and it may also contribute to the different result.
\begin{figure}
 \begin{minipage}[h]{0.47\linewidth}
\center{ \includegraphics[width=1\textwidth]{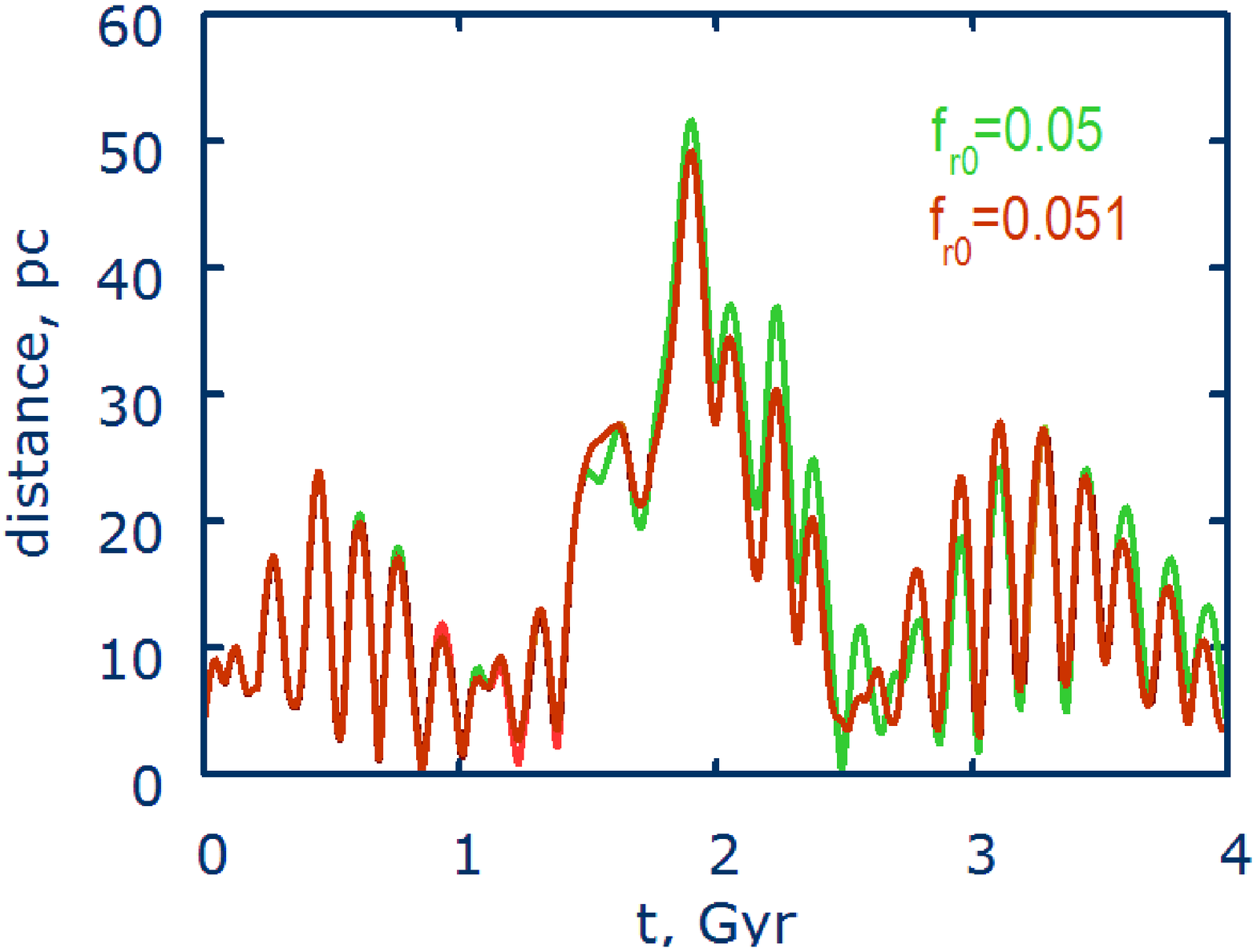} (a)}
\end{minipage}
\hfill
\begin{minipage}[h]{0.47\linewidth}
\center{\includegraphics[width=1\textwidth]{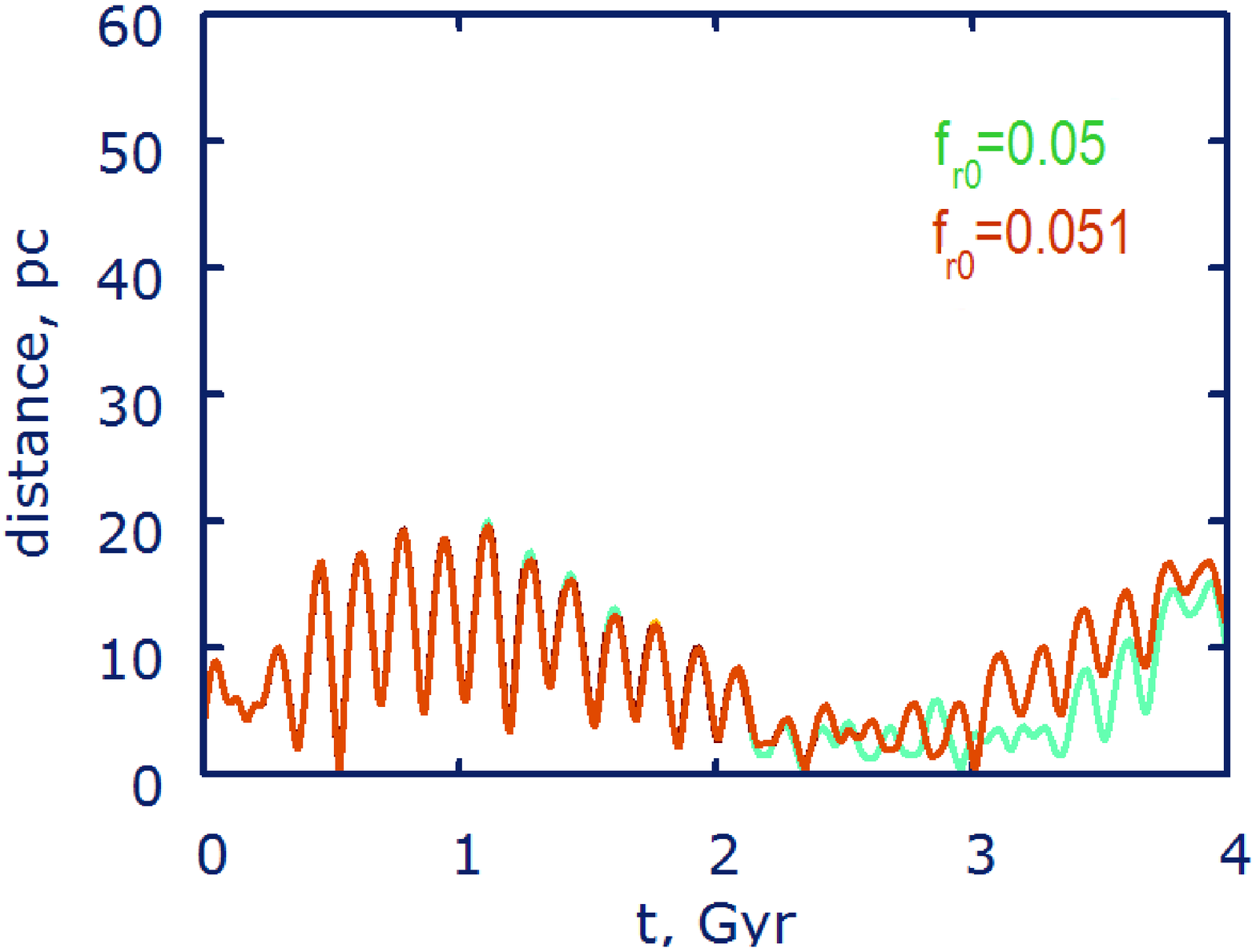} (b)}
\end{minipage}
\vfill
\begin{minipage}[h]{0.47\linewidth}
\center{\includegraphics[width=1\textwidth]{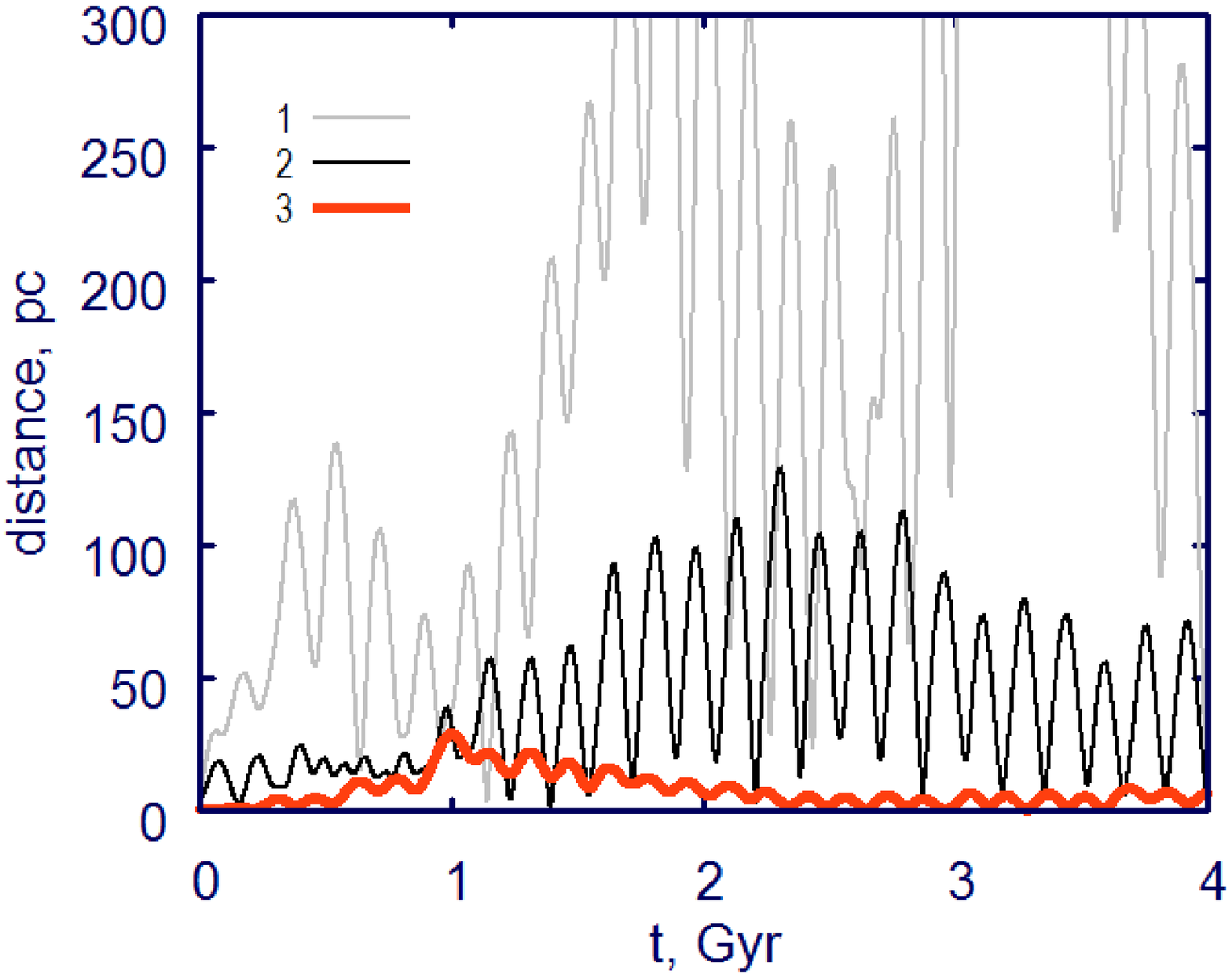} (c)}
\end{minipage}
\hfill
\begin{minipage}[h]{0.47\linewidth}
\center{\includegraphics[width=1\textwidth]{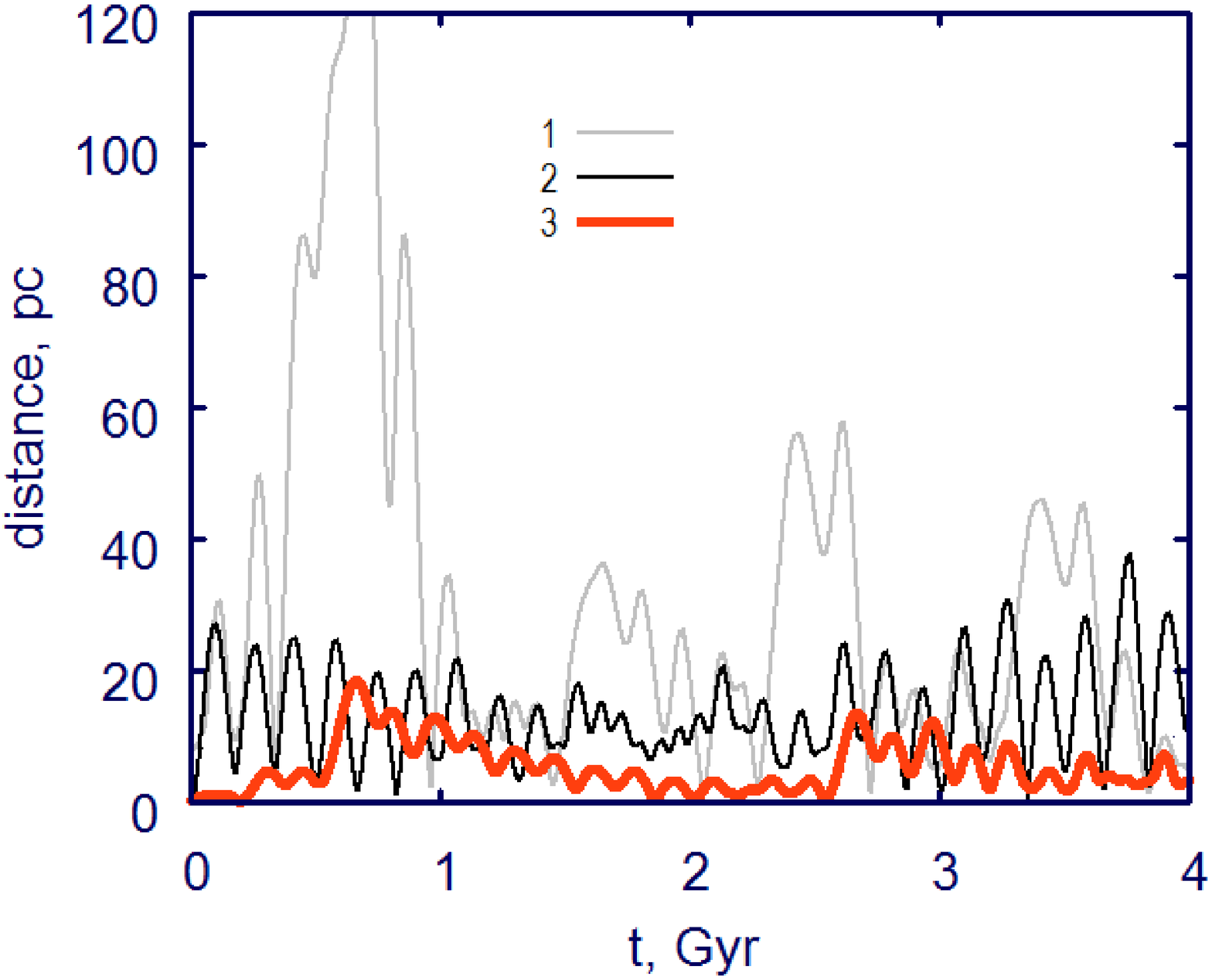} (d)}
\end{minipage}
\hfill
\caption{The separation of a star from the cluster center assuming the constant spiral strength model (a) or variable spiral strength (b). The two lines show the result for two slightly different values of the spiral potential strength. In panels (c) and (d) line 1 refers to Sellwood's (2010) $m=4$ model, line 2 to Gerhard's (2011) model with $m=4$ while line 3 refers to Sellwood's model with $m=2$. In panel (c) we use the variable spiral arm model, in panel (d) the constant arm model.}
\label{fig:3(a)}       
\end{figure}
%
%

 Therefore the search for Solar Siblings is a worth while task; there should be a reasonable number of them in our near Galactic neighborhood. However, our study also shows the sensitivity to Galactic models. It is advisable that the candidates for Solar Siblings should be studied in a range of Galactic Models to see how robust the orbital convergence between the Sun and the Sibling candidate is.


%
\begin{table}
\caption{The mean number of stars within distance R (in pc) from the cluster centre after 2 and 4 G yr in different models of constant or variable density wave obtained using 100 random realizations in a Monte-Carlo simulation where the spiral wave parameters are distributed over a wide range ($\sigma_i=nominal_i/2, f_{r0,nominal}=0.05$).}
$
\begin{tabular}{lrrrrrr}

\hline\\
                    &  R=100 & R=50  &  R=25    &           R=100 & R=50  &  R=25  \\

                   &  [pc]  & [pc]  &  [pc]  &          [pc] & [pc] & [pc] \\

                           &   &      2 G yr                 &    &                   &  4 G yr               \\

\hline\\
Sellwood:\\
m=4\\
 $\Omega_{p}=18\pm9$ km/s/kpc\\
$i=-13\pm6.5^{\circ}$      & 32     & 13     & 5                   &       16     &  6     & 2      \\

 $f_{r0}=0.05\pm0.025$ \\
 $\chi_\odot=-120\pm60^{\circ}$\\\
Gerhard:\\
m=4\\

 $\Omega_{p}=25\pm12.5$ km/s/kpc\\
$i=-13\pm6.5^{\circ}$    & 44     &  16     & 6                  &       16     & 6     & 2      \\

 $f_{r0}=0.05\pm0.025$\\
 $\chi_\odot=-120\pm60^{\circ}$\\\
Sellwood:\\
m=2\\
 $\Omega_{p}=8.1\pm4.05$ km/s/kpc\\
$i=-6\pm3^{\circ}$   & 53     & 22     & 8                   &        21     &  8     & 3      \\

 $f_{r0}=0.05\pm0.025$\\
 $\chi_\odot=-120\pm60^{\circ}$\\\
Sellwood:\\
m=4\\
 $\Omega_{p}=18\pm9$ km/s/kpc\\
$i=-13\pm6.5^{\circ}$    & 44     &  19     & 6                   &       18     &  7     & 2      \\

 $f_{r0}$=var(max $0.05\pm0.025$)\\
 $\chi_\odot=-120\pm60^{\circ}$\\\
\
Gerhard:\\
m=4\\
 $\Omega_{p}=25\pm12.5$ km/s/kpc\\
$i=-13\pm6.5^{\circ}$    & 38     &  17     & 6                 &          18     &  8     & 3      \\

 $f_{r0}$=var(max $0.05\pm0.025$)\\
 $\chi_\odot=-120\pm60^{\circ}$\\\
Sellwood:\\
m=2\\
 $\Omega_{p}=8.1\pm4.05$ km/s/kpc\\
$i=-6\pm3^{\circ}$   & 55     & 24     & 9                   &        33     &  15     & 6      \\

 $f_{r0}$=var(max $0.05\pm0.025$)\\
 $\chi_\odot=-120\pm60^{\circ}$\\
 
     \\

\hline\\

\end{tabular}
$
\end{table}

\begin{table}
\caption{The mean number of stars within distance R (in pc) from the cluster centre after 2 and 4 G yr in different models of constant or variable density wave obtained using 100 random realizations in a Monte-Carlo simulation where the spiral wave parameters are distributed over a wide range ($\sigma_i=nominal_i/2, f_{r0,nominal}=0.1$).}
$
\begin{tabular}{lrrrrrr}

\hline\\
                    &  R=100 & R=50  &  R=25    &           R=100 & R=50  &  R=25  \\

                   &  [pc]  & [pc]  &  [pc]  &          [pc] & [pc] & [pc] \\

                           &   &      2 G yr                 &    &                   &  4 G yr               \\

\hline\\
Sellwood:\\
m=4\\
 $\Omega_{p}=18\pm9$ km/s/kpc\\
$i=-13\pm6.5^{\circ}$      & 18     & 8     & 3                   &       7     &  3     & 1      \\

 $f_{r0}=0.1\pm0.05$ \\
 $\chi_\odot=-120\pm60^{\circ}$\\\
Gerhard:\\
m=4\\

 $\Omega_{p}=25\pm12.5$ km/s/kpc\\
$i=-13\pm6.5^{\circ}$    & 25     &  10     & 4                  &       10     & 4     & 1      \\

 $f_{r0}=0.1\pm0.05$\\
 $\chi_\odot=-120\pm60^{\circ}$\\\
Sellwood:\\
m=2\\
 $\Omega_{p}=8.1\pm4.05$ km/s/kpc\\
$i=-6\pm3^{\circ}$   & 23     & 9     & 3                   &        14     &  5     & 2      \\

 $f_{r0}=0.1\pm0.05$\\
 $\chi_\odot=-120\pm60^{\circ}$\\\
Sellwood:\\
m=4\\
 $\Omega_{p}=18\pm9$ km/s/kpc\\
 $i=-13\pm6.5^{\circ}$   & 19     &  8     & 3                   &       8     &  3     & 1      \\

 $f_{r0}$=var(max $0.1\pm0.05$)\\
 $\chi_\odot=-120\pm60^{\circ}$\\\
Gerhard:\\
m=4\\
 $\Omega_{p}=25\pm12.5$ km/s/kpc\\
$i=-13\pm6.5^{\circ}$    & 27     &  11     & 4                 &          10     &  4     & 1      \\

 $f_{r0}$=var(max $0.1\pm0.05$)\\
 $\chi_\odot=-120\pm60^{\circ}$\\\
Sellwood:\\
m=2\\
 $\Omega_{p}=8.1\pm4.05$ km/s/kpc\\
$i=-6\pm3^{\circ}$   & 31     & 12     & 4                   &        10     &  4     & 1      \\

 $f_{r0}$=var(max $0.1\pm0.05$)\\
  $\chi_\odot=-120\pm60^{\circ}$\\
     \\

\hline\\

\end{tabular}
$
\end{table}

\begin{table}
\caption{The mean number of stars within distance R (in pc) from the cluster centre after 2 and 4 G yr in different models of constant or variable density wave $f_{r0}=0.05,0.1$.}
$
\begin{tabular}{lrrrrrr}

\hline\\
                    &  R=100 & R=50  &  R=25    &           R=100 & R=50  &  R=25  \\

                   &  [pc]  & [pc]  &  [pc]  &          [pc] & [pc] & [pc] \\

                           &   &      2 G yr                 &    &                   &  4 G yr               \\

\hline\\
Sellwood:\\
m=4\\
 $\Omega_{p}=18$ km/s/kpc      & 105     & 61     & 27                   &       81     &  31     & 8      \\

 $f_{r0}=0.05$ \\\
Gerhard:\\
m=4\\

 $\Omega_{p}=25$ km/s/kpc    & 30     &  10     & 5                  &       16     & 8     & 4      \\

 $f_{r0}=0.05$\\\
Sellwood:\\
m=2\\
 $\Omega_{p}=8.1$ km/s/kpc   & 135     & 75     & 41                   &        66     &  24     & 7      \\

 $f_{r0}=0.05$\\\
Sellwood:\\
m=4\\
 $\Omega_{p}=18$ km/s/kpc    & 56     &  17     & 6                   &       25     &  8     & 1      \\

 $f_{r0}$=var(max 0.05)\\\
Gerhard:\\
m=4\\
 $\Omega_{p}=25$ km/s/kpc    & 22     &  4     & 1                 &          13     &  8     & 3      \\

 $f_{r0}$=var(max 0.05)\\\
Sellwood:\\
m=2\\
 $\Omega_{p}=8.1$ km/s/kpc   & 124     & 74     & 33                   &        62     &  32     & 19      \\

 $f_{r0}$=var(max 0.05)\\\
Sellwood:\\
m=4\\
 $\Omega_{p}=18$ km/s/kpc      & 0     & 0     & 0                   &       0     &  0     & 0      \\

 $f_{r0}=0.1$ \\\
Gerhard:\\
m=4\\

 $\Omega_{p}=25$ km/s/kpc    & 11     &  6     & 1                  &       13     & 2     & 1      \\

 $f_{r0}=0.1$\\\
Sellwood:\\
m=2\\
 $\Omega_{p}=8.1$ km/s/kpc   & 6     & 0     & 0                   &        3     &  0     & 0      \\

 $f_{r0}=0.1$\\\
Sellwood:\\
m=4\\
 $\Omega_{p}=18$ km/s/kpc    & 17     &  7     & 0                   &       0     &  0     & 0      \\

 $f_{r0}$=var(max 0.1)\\\
Gerhard:\\
m=4\\
 $\Omega_{p}=25$ km/s/kpc    & 1     &  0     & 0                 &          2     &  1     & 1      \\

 $f_{r0}$=var(max 0.1)\\\
Sellwood:\\
m=2\\
 $\Omega_{p}=8.1$ km/s/kpc   & 18     & 8     & 2                   &        16     &  9     & 4      \\

 $f_{r0}$=var(max 0.1)\\ 
     \\

\hline\\

\end{tabular}
$
\end{table}

\begin{table}
\caption{The mean number of stars within distance R (in pc) from the cluster centre after 2 and 4 G yr in different models of constant or variable density wave obtained using 100 random realizations in a Monte-Carlo simulation where $f_{r0}$=$0.1\pm0.05$, other spiral wave parameters are fixed.}
$
\begin{tabular}{lrrrrrr}

\hline\\
                    &  R=100 & R=50  &  R=25    &           R=100 & R=50  &  R=25  \\

                   &  [pc]  & [pc]  &  [pc]  &          [pc] & [pc] & [pc] \\

                           &   &      2 G yr                 &    &                   &  4 G yr               \\

\hline\\
Sellwood:\\
m=4\\
 $\Omega_{p}=18$ km/s/kpc\\
$i=-13^{\circ}$      & 21     & 9     & 3                   &       9     &  4     & 1      \\

 $f_{r0}=1.0\pm0.05$ \\
 $\chi_\odot=-120^{\circ}$\\\
Gerhard:\\
m=4\\

 $\Omega_{p}=25$ km/s/kpc\\
$i=-13^{\circ}$    & 22     &  9     & 4                  &       14     & 6     & 2      \\

 $f_{r0}=1.0\pm0.05$\\
 $\chi_\odot=-120^{\circ}$\\\
Sellwood:\\
m=2\\
 $\Omega_{p}=8.1$ km/s/kpc\\
$i=-6^{\circ}$   & 21     & 9     & 3                   &        9     &  3     & 1      \\

 $f_{r0}=1.0\pm0.05$\\
 $\chi_\odot=-120^{\circ}$\\\
Sellwood:\\
m=4\\
 $\Omega_{p}=18$ km/s/kpc\\
$i=-13^{\circ}$    & 51     &  22     & 8                   &       7     &  2     & 1      \\

 $f_{r0}$=var(max $0.1\pm0.05$)\\
 $\chi_\odot=-120^{\circ}$\\\
Gerhard:\\
m=4\\
 $\Omega_{p}=25$ km/s/kpc\\
$i=-13^{\circ}$    & 12     &  5     & 2                 &          4     &  2     & 1      \\

 $f_{r0}$=var(max $0.1\pm0.05$)\\
 $\chi_\odot=-120^{\circ}$\\\
Sellwood:\\
m=2\\
 $\Omega_{p}=8.1$ km/s/kpc\\
$i=-6^{\circ}$   & 26     & 11     & 4                   &        17     &  7     & 2      \\

 $f_{r0}$=var(max $0.1\pm0.05$)\\
 $\chi_\odot=-120^{\circ}$\\
 
     \\

\hline\\

\end{tabular}
$
\end{table}
\begin{acknowledgements}
This work was supported by the ``Nonstationary Phenomena in
Objects of the Universe'' Program of the Presidium of the Russian
Academy of Sciences and the ``Multiwavelength Astrophysical
Research'' grant no. NSh--16245.2012.2 from the President of the
Russian Federation. It is also supported by grants from Finnish Society for Sciences and Letters and from Finnish Academy of Science and Letters.

\end{acknowledgements}


\begin{thebibliography}{}
\bibitem{Adams2010}
Adams, F. C.,: The Birth Environment of the Solar System. Annual Review of Astronomy and Astrophysics {\bf 48}, 47-85 (2010)

\bibitem{Antoja2009}
Antoja, T., Valenzuela, O., Pichardo, B., Moreno, E., Figueras, F., Fernandez, D.: Stellar Kinematic Constraints on Galactic Structure Models Revisited: Bar and Spiral Arm Resonances. Astrophysical Journal Letters {\bf 700}, L78-L82 (2009)
	
\bibitem{Babaetal2009}
Baba, J., Asaki, Y., Makino, J., Miyoshi, M., Saitoh, T. R., Wada, K.: 	The Origin of Large Peculiar Motions of Star-Forming Regions and Spiral Structures of Our Galaxy. Astrophysical Journal {\bf 706}, 471-481 (2009)
	
\bibitem{BatistaandFernandes2012}
Batista, S. F. A. and Fernandes, J.: Lost siblings of the Sun: Revisiting the FGK potential candidates. New Astronomy {\bf 17}, 514-519 (2012)

\bibitem{BobylevandBajkova2010}
Bobylev, V. V. and Bajkova, A. T.: Galactic parameters from masers with trigonometric parallaxes. Monthly Notices of the Royal Astronomical Society {\bf 408}, 1788-1795 (2010)

\bibitem{BobylevandBajkova2013}
Bobylev, V. V. and Bajkova, A.T.: The Milky Way spiral structure parameters from data on masers and selected open lusters. Monthly Notices of the Royal Astronomical Society {\bf 437}, 1549 - 1553 (2014) 

\bibitem{Bobylevetal2011} 
Bobylev, V. V., Bajkova, A. T., Myll\"ari, A. and Valtonen, M.:  Searching for possible siblings of the sun from a common cluster based on stellar space velocities, Astronomy Letters {\bf 37}, 550 - 562 (2011)

\bibitem{Bobylevetal2008}
Bobylev, V. V., Bajkova, A. T. and Stepanishchev, A. S.: Galactic rotation curve and the effect of density waves from data on young objects. Astronomy Letters {\bf 34}, 515-528 (2008)

\bibitem{Brownetal2010}
Brown, A. G. A., Portegies Zwart, S. F. and Bean, J.: The quest for the Sun's siblings: an exploratory search in the Hipparcos Catalogue. Monthly Notices of the Royal Astronomical Society {\bf407}, 458-464 (2010)

\bibitem{Englmaieretal2008}
Englmaier, P., Pohl, M. and Bissantz, N.: The Milky Way Spiral Arm Pattern. arXiv:0812.3491 (2008)

\bibitem{Fellhaueretal2006}
Fellhauer, M., Belokurov, V., Evans, N. W., Wilkinson, M. I., Zucker, D. B., Gilmore, G., Irwin, M. J., Bramich, D. M., Vidrih, S., Wyse, R. F. G., Beers, T. C., and Brinkmann, J.: The Origin of the Bifurcation in the Sagittarius Stream. Astrophysical Journal {\bf 651}, 167-173 (2006)

\bibitem{Fernandezetal2001}
Fernandez, D., Figueras, F. and Torra, J.: Kinematics of young stars. II. Galactic spiral structure. Astronomy and Astrophysics {\bf 372}, 833-850 (2001)

\bibitem{Fernandezetal2008}
Fernandez, D., Figueras, F. and Torra, J.: On the kinematic evolution of young local associations and the Scorpius-Centaurus complex. Astronomy and Astrophysics {\bf 480}, 735-751 (2008)

\bibitem{Gerhard2011}
Gerhard, O.: Pattern speeds in the Milky Way. Memorie della Societa Astronomica Italiana Supplement {\bf 18}, 185-188 (2011)

\bibitem{Hahnetal2011}
Hahn, C. H., Sellwood, J. A. and Pryor, C.: Velocity-space substructure from nearby RAVE and SDSS stars. Monthly Notices of the Royal Astronomical Society {\bf 418}, 2459-2466 (2011)

\bibitem{Hernquist1990}
Hernquist, L.: An analytical model for spherical galaxies and bulges. Astrophysical Journal {\bf 356}, 359-364 (1990)

\bibitem{Houetal2009}
Hou, L. G., Han, J. L. and Shi, W. B.: The spiral structure of our Milky Way Galaxy. Astronomy and Astrophysics {\bf 499}, 473-482 (2009)

	
\bibitem{Joshi2007}
Joshi, Y. C.: Displacement of the Sun from the Galactic plane. 	Monthly Notices of the Royal Astronomical Society {\bf 378}, 768-776 (2007)

\bibitem{Lepineetal2001}
Lepine, J. R. D., Mishurov, Yu. N. and Dedikov, S. Yu.: A New Model for the Spiral Structure of the Galaxy: Superposition of 2- and 4-armed Patterns. The Astrophysical Journal {\bf 546}, 234-247 (2001)

\bibitem{LinandShu1964}
Lin, C. C. and Shu, F. H.: On the Spiral Structure of Disk Galaxies. Astrophysical Journal {\bf 140}, 646-655 (1964)

\bibitem{Linetal1969}
Lin, C. C., Yuan, C. and Shu, F. H.: On the Spiral Structure of Disk Galaxies. III. Comparison with Observations. Astrophysical Journal {\bf 155}, 721-746 (1969)

\bibitem{MelnikandRautiainen2009}
Mel'nik, A. M. and Rautiainen, P.: Kinematics of the outer pseudorings and the spiral structure of the Galaxy. Astronomy Letters {\bf 35}, 609-624 (2009)


\bibitem{MishurovandAcharova2011}
Mishurov, Yu. N. and Acharova, I. A.: Is it possible to reveal the lost siblings of the Sun? Monthly Notices of the Royal Astronomical Society {\bf 412}, 1771-1777 (2011)

\bibitem{MishurovandZenina1999}
Mishurov, Yu. N. and Zenina, I. A.: Yes, the Sun is located near the corotation circle. Astronomy and Astrophysics {\bf 341}, 81-85 (1999)

\bibitem{MiyamotoandNagai1975}
Miyamoto, M. and Nagai, R.: Three-dimensional models for the distribution of mass in galaxies. Publications of Astronomical Society of Japan {\bf 27}, 533-543 (1975)



\bibitem{NaozandShaviv2007}
Naoz, S. and Shaviv, N. J.: Open cluster birth analysis and multiple spiral arm sets in the Milky Way. New Astronomy {\bf 12}, 410-421 (2007)


\bibitem{PopovaandLoktin2005}
Popova, M. E. and Loktin, A. V.: Parameters of the Spiral Structure of the Galaxy from Data on Open Star Clusters. Astronomy Letters {\bf 31}, 171-178 (2005)

\bibitem{PortegiesZwart2009}
Portegies Zwart, S.F.: The Lost Siblings of the Sun. Astrophysical Journal Letters {\bf 696}, L13-L16 (2009)

\bibitem{Ramirez2014}
Ramirez, I.; Bajkova, A. T.; Bobylev, V. V.; Roederer, I. U.; Lambert, D. L.; Endl, M.; Cochran, W. D.; MacQueen, P. J.; Wittenmyer, R. A.: Elemental Abundances of Solar Sibling Candidates. Astrophysical Journal {\bf 787}, article id. 154 (2014)

\bibitem{Russeil2003}
Russeil, D.: Star-forming complexes and the spiral structure of our Galaxy. Astronomy and Astrophysics {\bf 397}, 133-146 (2003)

\bibitem{Sellwood2010}
Sellwood, J. A.: A recent Lindblad resonance in the solar neighbourhood. 	Monthly Notices of the Royal Astronomical Society {\bf 409}, 145-155 (2010)


\bibitem{Sellwood2012}
Sellwood, J. A.: Spiral Instabilities in N-body Simulations. I. Emergence from Noise. Astrophysical Journal {\bf 751}, 44-54 (2012)

\bibitem{Sellwood2013}
Sellwood, J. A.: Secular Evolution in Disk Galaxies. Reviews of Modern Physics (in press, arXiv:1310.0403, 2013) 

\bibitem{SellwoodandBinney2002}
Sellwood, J. A. and Binney, J. J.: Radial mixing in galactic discs. Monthly Notice of the Royal Astronomical Society {\bf336}, 785-796 (2002)

\bibitem{Schonrichetal2010}
Sch\"onrich, R., Binney, J. and Dehnen, W.: Local kinematics and the local standard of rest. Monthly Notices of the Royal Astronomical Society {\bf 403}, 1829-1833 (2010)

	
\bibitem{Valtonenetal2009}
Valtonen, M., Nurmi, P., Zheng, J.-Q., Cucinotta, F. A., Wilson, J. W., Horneck, G., Lindegren, L., Melosh, J., Rickman, H. and Mileikowsky, C.:  Natural Transfer of Viable Microbes in Space from Planets in Extra-Solar Systems to a Planet in our Solar System and Vice Versa. Astrophysical Journal, {\bf 690}, 210 - 215 (2009)

\bibitem{Wielenetal1996}
Wielen, R., Fuchs, B. and Dettbarn, C.: On the birth-place of the Sun and the places of formation of other nearby stars. Astronomy and Astrophysics {\bf 314}, 438-447 (1996)

\bibitem{Yuan1969}
Yuan, C.: Application of the Density-Wave Theory to the Spiral Structure of the Milky way System. II. Migration of Stars. Astrophysical Journal {\bf 158}, 889-898 (1969)





%
%



\end{thebibliography}


\end{document}